\documentclass[usenatbib]{mnras}

\usepackage{newtxtext,newtxmath}

\usepackage[T1]{fontenc}
\usepackage{ae,aecompl}

\usepackage{mathtools}
\usepackage{siunitx}
\usepackage{amssymb,multirow,tabularx}
\usepackage{amsmath}
\usepackage{graphicx}
\usepackage{comment}
\usepackage[normalem]{ulem}
\usepackage{BibDef}
\usepackage{xcolor}
\usepackage{hyperref}
\hypersetup{
	colorlinks=true,        
	linkcolor=blue,         
	citecolor=blue,         
}

\newcommand{\msun}{~\rm M_{\large \odot}}
\newcommand{\ud}{~{\rm d}}

\definecolor{green2}{RGB}{21, 158, 14}

\title[Eccentric MBHB evolution]{On the eccentricity evolution of massive black hole binaries in stellar backgrounds}

\author[M. Bonetti et al.]{Matteo Bonetti$^{1,2}$, Alexander Rasskazov$^3$, Alberto Sesana$^{1,2}$, Massimo Dotti$^{1,2}$, 
\newauthor Francesco Haardt$^{2,4}$, Nathan W. C. Leigh$^{5,6}$, Manuel Arca Sedda$^{7}$, Giacomo Fragione$^{8,9}$ 
\newauthor and Elena Rossi$^{10}$\\
$^1$Dipartimento di Fisica ``G. Occhialini'', Universit\`a degli Studi di Milano-Bicocca, Piazza della Scienza 3, 20126 Milano, Italy\\
$^2$INFN, Sezione di Milano-Bicocca, Piazza della Scienza 3, 20126 Milano, Italy\\
$^3$E\"otv\"os University, Institute of Physics, P\'azm\'any P. s. 1/A, Budapest, Hungary 1117\\
$^4$Dipartimento di Scienza e Alta Tecnologia,  Universit\`a degli Studi dell'Insubria, Via Valleggio 11, 22100, Como, Italy\\
$^5$Departamento de Astronom\'ia, Facultad de Ciencias F\'isicas y Matem\'aticas, Universidad de Concepci\'on, Chile\\
$^6$Department of Astrophysics, American Museum of Natural History, Central Park West and 79th Street, New York, NY 10024\\
$^7$Astronomisches Rechen Institut, Zentrum f\"ur Astronomie der Universit\"at Heidelberg, Monchhofstrasse 12-14, 69120 Heidelberg, Germany \\
$^8$Department of Physics \& Astronomy, Northwestern University, Evanston, IL 60202, USA\\
$^9$Center for Interdisciplinary Exploration \& Research in Astrophysics (CIERA), Evanston, IL 60202, USA\\
$^{10}$Leiden Observatory, Leiden University, PO Box 9513, 2300 RA, Leiden, the Netherlands\\
}

\date{Accepted 2020 February 03. Received 2020 January 31; in original form 2019 December 24}

\pubyear{2020}

\begin{document}
\label{firstpage}
\pagerange{\pageref{firstpage}--\pageref{lastpage}}
\maketitle

\begin{abstract}
  We study the dynamical evolution of eccentric massive black hole binaries (MBHBs) interacting with unbound stars by means of an extensive set of three body scattering experiments. Compared to previous studies, we extend the investigation down to a MBHB mass ratio of $q=m_2/m_1=10^{-4}$, where $m_1$ and $m_2$ are the masses of the primary and secondary hole respectively. Contrary to a simple extrapolation from higher mass ratios, we find that for $q\lesssim 10^{-3}$ the eccentricity growth rate becomes negative, i.e., the binary {\it circularises} as it shrinks. This behaviour is due to the subset of interacting stars captured in metastable counter-rotating orbits; those stars tend to extract angular momentum from the binary, promoting eccentricity growth for $q>10^{-3}$, but tend to inject angular momentum into the binary driving it towards circularisation for $q<10^{-3}$. The physical origin of this behaviour requires a detailed study of the orbits of this subset of stars and is currently under investigation. Our findings might have important consequences for intermediate MBHs (IMBHs) inspiralling onto MBHs (e.g. a putative $10^3\msun$ black hole inspiralling onto SgrA$^*$).
\end{abstract}

\begin{keywords}
galaxies: nuclei -- stars: kinematics and dynamics -- black hole physics -- gravitational waves
\end{keywords}

\section{Introduction}

Massive black holes (MBHs) are among the fundamental building blocks of cosmic structures \citep[e.g.][ and references therein]{2006ApJS..163....1H,2013ARA&A..51..511K}. Residing in the centre of galaxies, they are surrounded by a dense gaseous and stellar environment, which naturally promotes strong dynamical interactions with other massive objects. This is mostly because dynamical friction (DF) tends to bring massive objects together at the centre of dense systems \citep{1943ApJ....97..255C}. Examples of this mechanism are the pairing of two MBHs into a binary (MBHB) following a galaxy merger \citep{1980Natur.287..307B}, or the inspiral of dense stellar clusters onto a galactic nucleus \citep{2008MNRAS.388L..69C}. In the latter case, if the cluster contains an intermediate MBH (IMBH) \citep{2002MNRAS.330..232C,2002ApJ...576..899P,gf2018a,gf2018b}, the eventual tidal disruption of the cluster will leave behind an IMBH-MBH binary \citep{2006ApJ...641..319P,2018MNRAS.477.4423A}. MBHs with mass $\sim10^6-10^7\msun$, residing in Milky Way-type galaxies, are therefore expected to form binaries with companions as light as $\sim 10^2$-$10^3\msun$, thus covering a range of mass ratios down to $q=m_2/m_1 = 10^{-4}$ (where $m_1$ is the mass of the primary hole). Understanding the dynamical evolution of these binaries is of paramount importance as they are anticipated to be the loudest gravitational wave (GW) sources for the Laser Interferometer Space Antenna \citep[LISA][]{2017arXiv170200786A}.

The dynamics of MBHBs in stellar environments have been studied by means of either semi-analytic models based on scattering experiments or full N-body simulations. Scattering experiments of unbound stars on MBHBs down to $q\sim0.004$ \citep{Quinlan1996,Sesana2006,Rasskazov2017} predict that the binary eccentricity in non-rotating stellar environments grows as the binary shrinks, which has been verified in N-body simulations of comparable mass binaries \citep[i.e. down to $q\sim0.1$, see e.g.,][]{2011ApJ...732L..26P,2012ApJ...749..147K,2012ApJ...744...74G,2014MNRAS.444...29L}. Lower mass ratios, down to $q=10^{-4}$, have been recently investigated by \citet[][R19 hereinafter]{Rasskazov2019} in the context of hypervelocity star production in the Galactic centre. They report an unexpected feature in the MBHB evolution for low mass ratios; at $q<10^{-3}$, three-body interactions appear to circularize the binary, contrary to the higher mass ratio cases. Since the GW decay time of circular binaries is longer \citep{1963PhRv..131..435P}, this finding can affect the overall merger rate of IMBH-MBH binaries in the Universe, promoting the formation of multiple IMBH systems orbiting a MBH, with interesting dynamical consequences.

In this letter, we carry an extensive series of three-body scattering experiments of initially unbound stars onto an eccentric MBHB with $e=0.6$ and $10^{-4}\leq q \leq 1$. The setup of the experiments is described in Section \ref{sec:method}. In Section \ref{sec:HK}, we report hardening and eccentricity growth rate as a function of $q$ confirming that, contrary to naive expectations, the MBHB tends to circularize for $q<10^{-3}$. We dissect the energy and angular momentum exchange between the MBHB and the incoming stars in Section \ref{sec:HKdissect}, identifying the counter-rotating stars captured in metastable orbits as the subpopulation responsible for the somewhat unexpected behaviour of the MBHB. Finally, we discuss our results and future work in Section \ref{sec:discussion}. 

\section{Scattering experiments: the method}
\label{sec:method}

Due to the lack of a general analytical solution to the three-body problem, any systematic investigation on this topic necessarily leverages on the numerical integration of the equations of motion. Employing a C++ implementation of the Bulirsch-Stoer \citep{Bulirsch1966} algorithm to evolve in time the three-body Newtonian equations of motion,\footnote{See \citet{Bonetti2016} for additional details about the code implementation. We stress that in the present work we do not consider general relativity corrections (through Post-Newtonian formalism), since on the scales typical for stellar hardening they are usually too weak to play an influential role.} we perform three-body scattering experiments of a MBHB interacting with a star coming from infinity with positive energy (i.e. unbound to the binary). We consider several MBHB mass ratios in the range $q \in [10^{-4},1]$, while fixing $m_1 = 10^6\msun$, the binary eccentricity to $e=0.6$ and the stellar intruder mass to $m_3 = 1\msun$. Simulation units are such that the initial binary semi-major axis is fixed to $a=1$ and the velocity of the intruder is rescaled accordingly, following the procedure employed in \citet{Sesana2006} and R19. The initial conditions, appropriate for a spherical, isotropic distribution, are set as follows:

\begin{itemize}
    \item the binary is initialised in the $x-y$ plane with pericentre along the positive $x$ axis; \footnote{This means that when the binary is at pericentre $m_1$ lies along positive $x$ axis, $m_2$ along the negative $x$ axis.}
    \item the velocity at infinity of the incoming star, $v$, is sampled with 80 log-uniform values in the range $3\times 10^{-3} \sqrt{\frac{q}{1+q}} < v/v_{\rm bin} < 30 \sqrt{\frac{q}{1+q}}$, with $v_{\rm bin}^2 = G M/a$, being $M = m_1+m_2$ (see Section~\ref{sec:HK} for the motivation of such choice);
    \item the impact parameter, $b$, is sampled assuming $b^2$ uniformly distributed in the range $\left[0,25\left(1+\frac{2GM}{5 v^2}\right)\right]$, i.e. the pericentre $r_p$ lies between 0 and $5a$;
    \item all angular variables are uniformly sampled in $[0,2\pi]$, except for the inclination which is chosen such that its cosine is uniformly distributed in $[-1,1]$.
\end{itemize}

For each value of $v$ we perform $5\times 10^4$ runs, for a grand total of $4\times 10^6$ simulations for each mass ratio. Numerical integration starts with the incoming star at separation of $50~a$ from the MBHB centre of mass (CoM). Following R19, we consider three similar stopping criteria:

\begin{itemize}
    \item[1.] the star leaves the sphere with radius $50 a$ with positive total energy;
    \item[2.] the final time of the three-body integration, $T_f$, exceeds a maximum allowed value, dependent on the binary properties, but always larger than the Hubble time at $z=0$;
    \item[3.] the total time spent inside the $50a$-sphere exceeds $2 \times 10^4$ binary orbital periods.
\end{itemize}
Every time the star leaves the $50a$-sphere with negative energy, we stop the integration and we collapse the MBHB into a single object of mass $m_1+m_2$. We then evolve the system star-MBHB assuming two-body Keplerian motion until the star re-enters the sphere, updating the MBHB orbital phase to the corresponding re-entrance time and finally resuming the three-body integration.\footnote{During the three-body direct integration energy and angular momentum are conserved at the level of machine precision ($10^{-16}$), with only a subdominant sample of simulations showing larger numerical errors up to $\Delta E_{\rm tot}/E_{\rm tot}, \Delta L_{\rm tot}/L_{\rm tot} \approx 10^{-9}$. Still this is much lower than the actual $E_\star$ (or $L_\star$) fractional change due to three-body interactions which is of the order of $m_3/M\sim 10^{-6}$.} In the following analysis we consider only ``resolved scatterings", i.e., those matching condition 1, which amounts to more than 99\% of the cases considered.

\section{Hardening and eccentricity growth rates: $H$ and $K$}
\label{sec:HK}

Considering a fixed isotropic spherically symmetric non-rotating stellar background with density $\rho$ and velocity dispersion $\sigma$, the evolution of a MBHB can be described through the determination of two dimensionless quantities, i.e., the hardening rate $H$ and the eccentricity growth rate $K$ \citep{Quinlan1996} defined as

\begin{align}
    H &= \dfrac{\sigma}{G \rho} \dfrac{\ud}{{\rm dt}}\left(\dfrac{1}{a}\right), \\
    K &= \dfrac{\ud e}{\ud\ln (1/a)}.
\end{align}
$H$ and $K$ are computed as a function of the binary hardness $a/a_h$, where $a_h = G m_2/ (4\sigma^2)$ is the hardening radius. This motivates the $q$-dependent choice of $v$ in Section \ref{sec:method}. Since we are mostly interested in binaries with $0.01<a/a_h<1$, and $a=1$ in simulation units, a MBHB with a semi-major axis of $a/a_h$ is then simulated by weighting the outcome of the experiments over a Maxwellian distribution with $\sigma=\sqrt{q/(1+q)}\sqrt{a/a_h}$. The chosen range for $v$ ensures that the Maxwellian distribution is well sampled down to the tails in the whole $a/a_h$ range of interest. More details about this procedure can be found in \citet{Quinlan1996} and \citet{Sesana2006}. 

\begin{figure}
    \centering
    \includegraphics[width=0.5\textwidth]{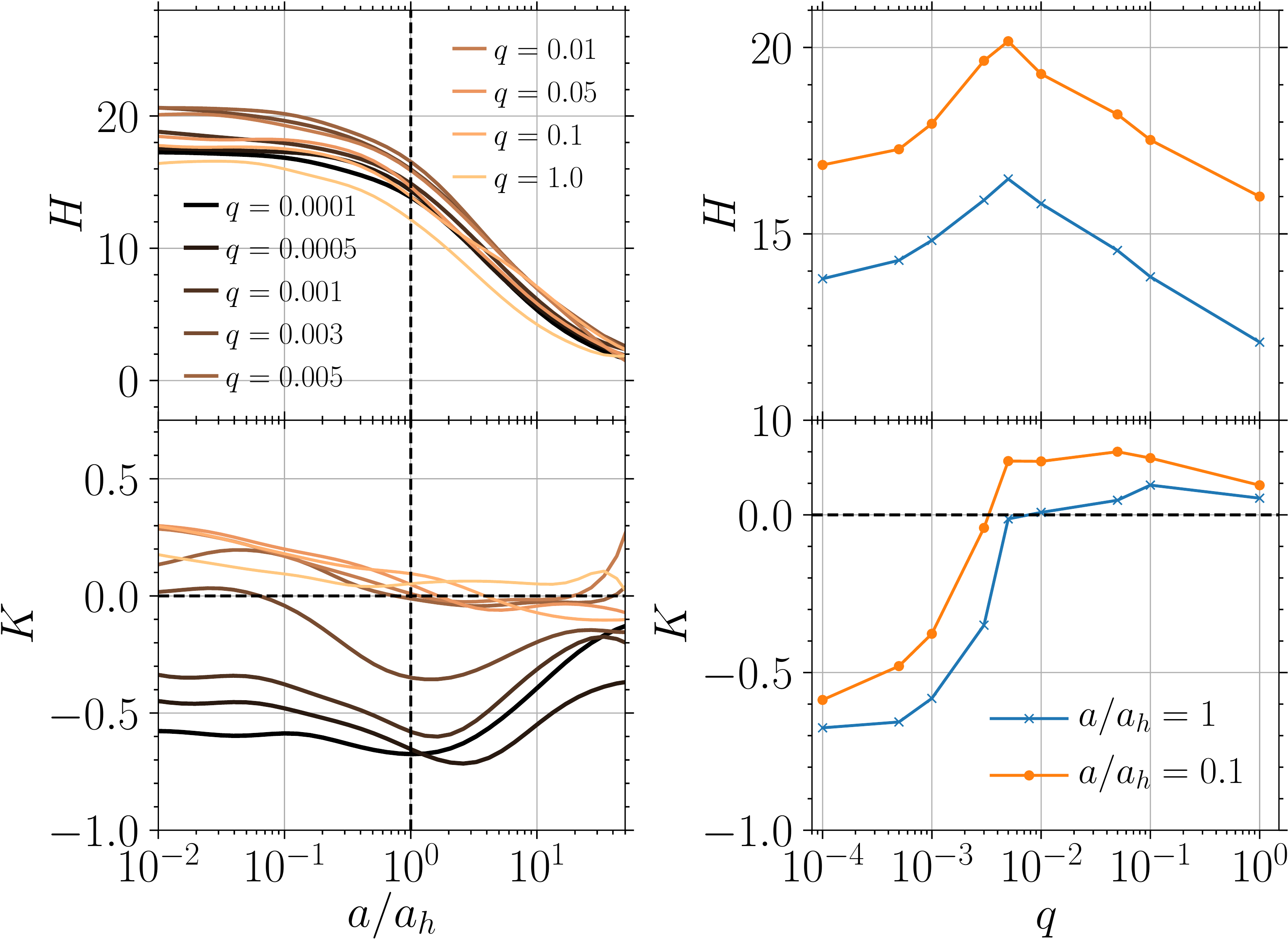}
    \caption{{\it Left panels}: hardening and eccentricity growth rates as a function of the binary hardness for several mass ratios as labelled. {\it Right panels}: values of $H$ and $K$ as a function of $q$ for fixed values of binary hardness. Note that below $q=0.005$ the eccentricity growth rate drops to negative values.}
    \label{fig:HK_comb}
\end{figure}
\begin{figure}
    \centering
    \includegraphics[width=0.47\textwidth]{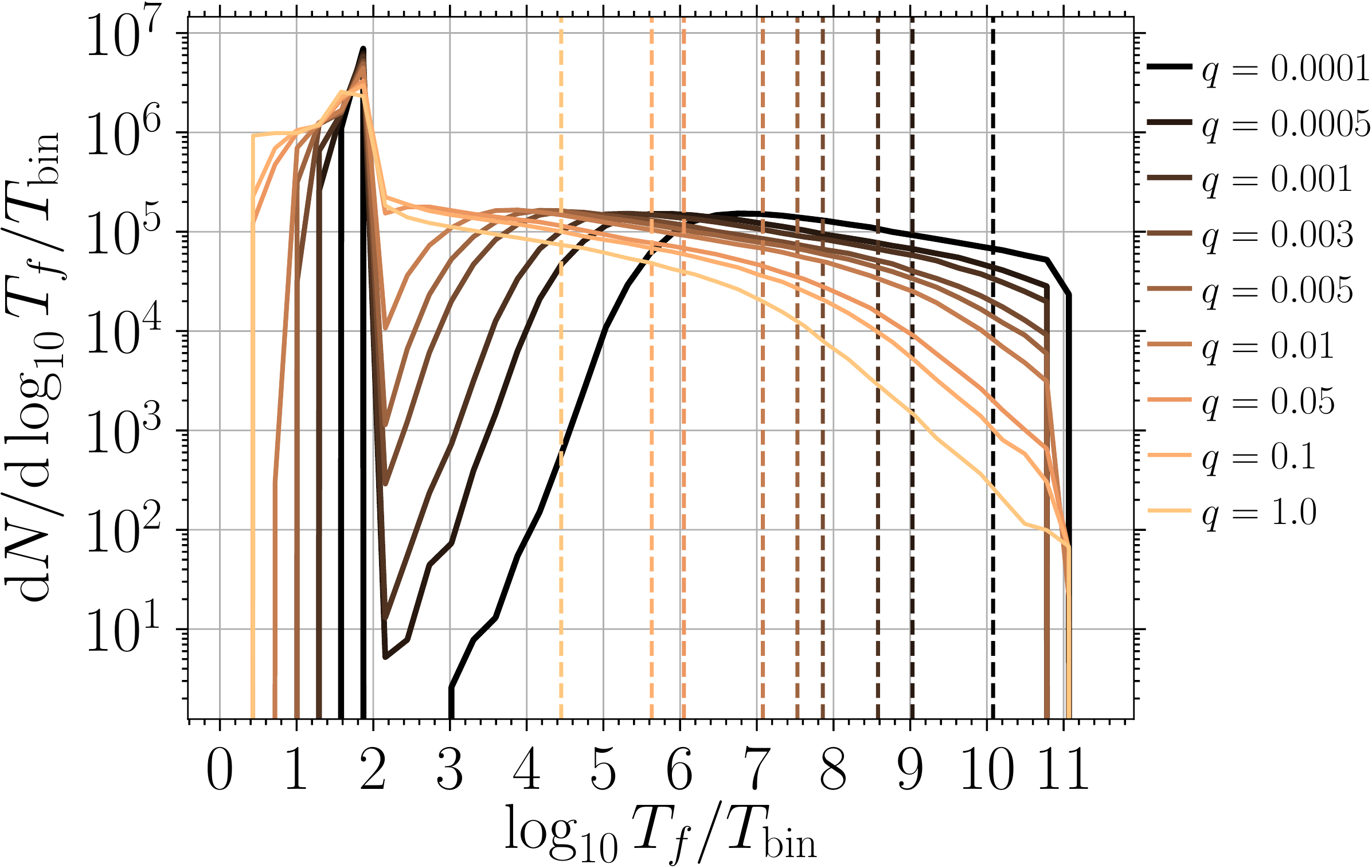}
    \caption{Distribution of the final time of the orbit integration in units of the binary period for different mass ratios. Vertical lines mark the corresponding $T=10^8$~yrs, the typical evolution timescale of the binary. Color-code as in Fig. \ref{fig:HK_comb}.}
    \label{fig:Tdist}
\end{figure}

Results for $H$ and $K$ as a function of $a/a_h$ and $q$ are shown in Fig.~\ref{fig:HK_comb}. The hardening rate is consistent with previous findings in the literature, with $H$ ramping-up as the binary shrinks and flattening to a value $15\lesssim H \lesssim 20$ for hard binaries (upper left panel). \textit{Note that $H$ tends to increase for decreasing $q$, but it turns over around $q\approx 0.005$.} Incidentally, this is roughly the minimum $q$ values at which $H$ has been computed for unbound scatterings to date ($q=1/256$ in \citealt{Quinlan1996} and $q=1/243$ in \citealt{Sesana2006}), so that this turnover has not been appreciated before.
 
More interesting is the behaviour of $K$, shown in the bottom panels. So long as $q>0.005$, $K\approx 0$ at large separation and progressively becomes positive as the binary hardens, consistent with the classic findings of \citet{Quinlan1996} and \citet{Sesana2006}. However, confirming what was previously noted by R19, $K$ plummets, becoming negative for $q\lesssim 0.003$. This is particularly evident in the bottom right panel of Fig.~\ref{fig:HK_comb}, which shows an essentially constant $K$ for $q\gtrsim 0.003$, and a sudden drop to large negative values for lower $q$. It is worth noticing two facts. First, the drop corresponds to the same value of $q$ for which the trend of $H$ with $q$ is reversed (see upper right panel), suggesting a likely correlation between the two. Second, it is not clear whether $K$ tends to zero for $a\gg a_h$ when $q < 0.005$ (lower left panel). There seems to be a tendency towards zero for $a>a_h$, but $K$ is still negative at $a/a_h>10$.

\subsection{Dissecting $H$ and $K$}
\label{sec:HKdissect}

An inspection of the interaction time distribution reveals that, as $q$ decreases, two distinct subsets of scatterings emerge, as shown in Fig. \ref{fig:Tdist} where we show the distribution of the final integration time $T_f$. We label ``quick-ejections'' those interactions in which the star returns to infinity after a single pericentre passage, and ``late-ejections'' those in which the star is captured by the MBHB into a loosely bound orbit. When $q\rightarrow 1$, late-ejections form a long tail in the interaction time distribution; the forcing quadrupolar potential is so strong that captures typically last only a few orbits. However for $q\ll 1$ the two populations are clearly distinct and the $K$ behaviour might be influenced by the relative weight of the two sub-populations. 
A second obvious way to separate the sample of interacting stars is in retrograde versus prograde orbits. A prograde orbit is defined by the condition ${\bf L_{\rm MBHB}\cdot L_*>0}$, where ${\bf L_{\rm MBHB}}$ is the orbital angular momentum vector of the MBHB and ${\bf L_*}$ is the initial angular momentum of the star with respect to the CoM of the MBHB. It has been shown both in bound  \citep{Sesana2011} and unbound \citep{Rasskazov2017} scattering experiments that initially prograde orbits tend to circularize the binary, whereas retrograde orbits drive its eccentricity upward. These results find confirmation in full $N$-body simulations \citep[see e.g.][]{2019MNRAS.484..520A}.
We thus divide the sample of interacting stars in four sub-populations: {\it quick-prograde, quick-retrograde, late-prograde, late-retrograde.}

\begin{figure}
    \centering
    \includegraphics[width=0.44\textwidth]{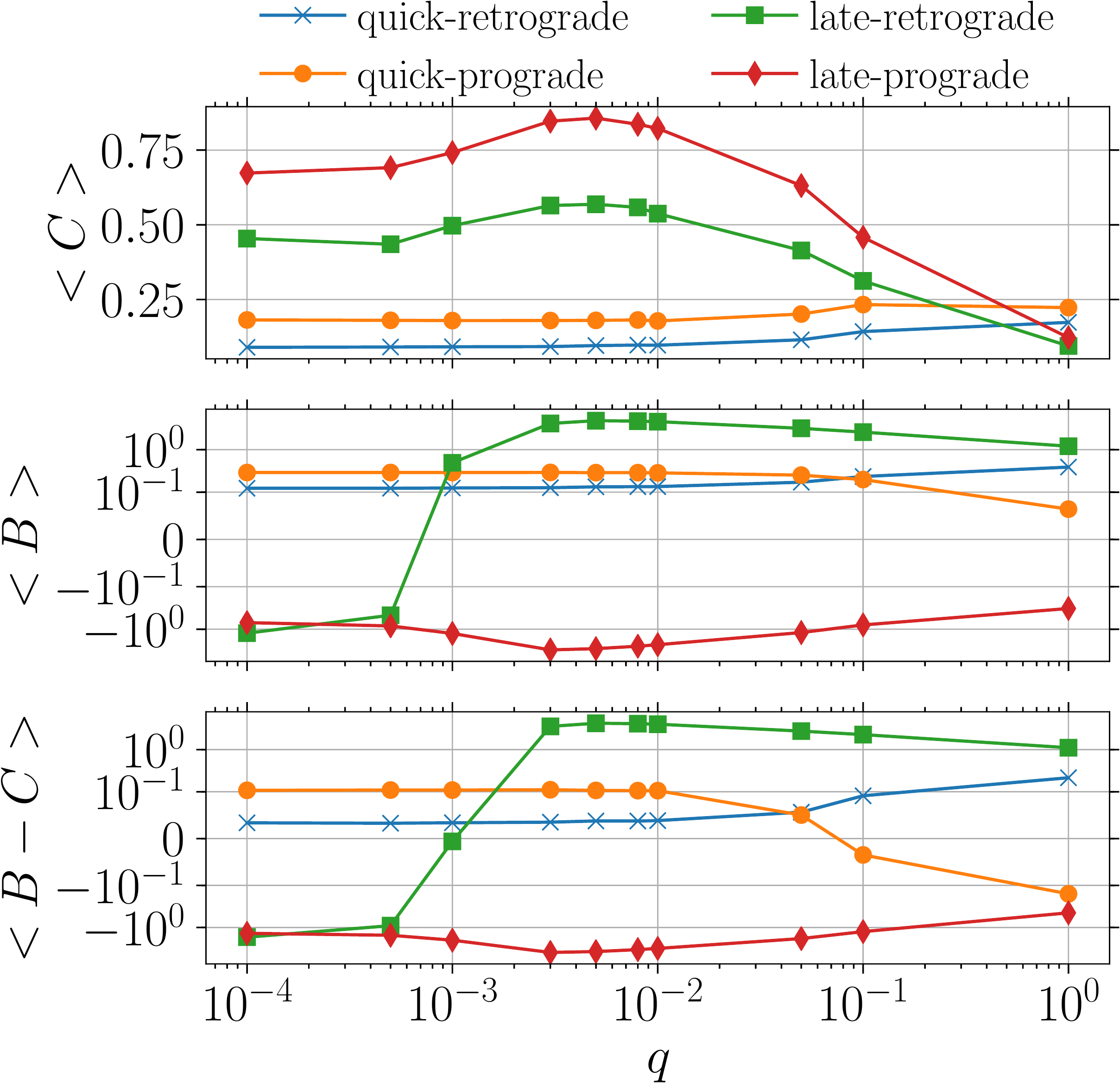}
    \caption{Mean value of $C$ (upper panel), $B$ (central panel) and $B-C$ (lower panel) when splitting simulations among quick/late and prograde/retrograde ejections as labelled.}
    \label{fig:C_B_mean}
\end{figure}
\begin{figure}
    \centering
    \includegraphics[width=0.44\textwidth]{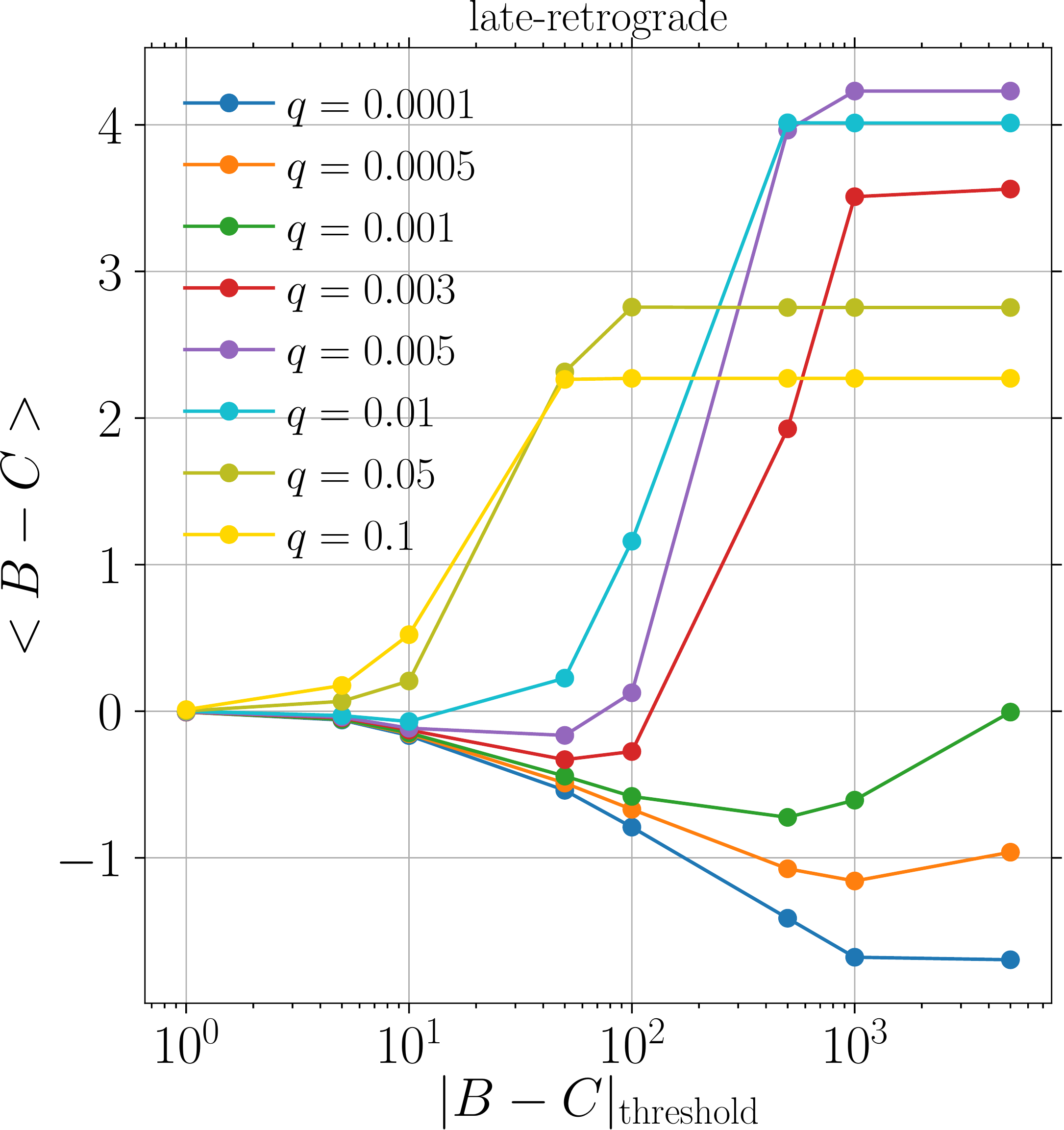}
    \caption{Mean value of $B-C$ evaluated as a function of an increasingly high threshold $|B-C|_{\rm threshold}$. Only late-retrograde ejections are considered.}
    \label{fig:BmC_mean_TH}
\end{figure}

In order to make our investigations cleaner we consider quantities that can be defined for every single star-MBHB interaction, instead of using $H$ and $K$, that are instead weighted over a Maxwellian velocity distribution. To this end, for each sub-population, we compute the average values of the dimensionless energy ($C$) and angular momentum ($B$) exchanged during the encounter, defined as:

\begin{align}
    C &= \dfrac{a \Delta E_{\star}}{G \mu \ m_3}\\
    B &= \dfrac{M}{\mu} \dfrac{\Delta L_{\star}}{h_{\rm bin} \ m_3},
\end{align}
where $h_{\rm bin} = L_{\rm bin}/\mu = \sqrt{G M a (1-e^2)}$ is the angular momentum per unit mass of the MBHB (being $\mu$ its reduced mass), while $\Delta E_\star$ and $\Delta L_\star$ are the stellar energy and angular momentum changes, respectively. In the above equations it is implicitly assumed that the total energy and angular momentum are conserved, such that any change in the MBHB energy/angular momentum is the negative of that of the star.
Finally, since the eccentricity change involves both energy and angular momentum variations, we also evaluate the difference $B-C$. It is straightforward to show that this quantity directly relates to $\Delta e$ through
\begin{equation}
  \Delta e = \dfrac{(1-e^2)}{e} \dfrac{m_3}{M} (B-C).
\label{eq:deltae}
\end{equation}

Results are shown in Fig.~\ref{fig:C_B_mean}, note that $\langle C\rangle$ and  $\langle B\rangle$ are defined as the energy and angular momentum changes {\it of the intruding star}. In the top panel, $\langle C\rangle$ shows that for $q<1$ late interactions are much more effective at extracting energy from the binary compared to quick ones. Moreover, within each sub-sample, prograde stars extract more energy than retrograde ones. Note the (broad) peak at $q = 0.005$, which can explain the turnover in the upper right panel of Fig.~\ref{fig:HK_comb}. As for $\langle B\rangle$, shown in the central panel of Fig.~\ref{fig:C_B_mean}, quick encounters (prograde and retrograde alike) tend to extract angular momentum from the MBHB, whereas things are more complicated for late encounters. For the latter, prograde orbits always tend to inject angular momentum into the MBHB, whereas retrograde orbits with $q>10^{-3}$ tend to extract it \citep[as discussed in][]{Sesana2011}. However, for $q<10^{-3}$ late-retrograde interactions, unexpectedly, also inject angular momentum into the MBHB. This is also more evident in the lower panel of Fig.~\ref{fig:C_B_mean}, where $\langle B-C\rangle$ (directly related to $\Delta e$) suddenly changes sign (green squares). In the same figure a similar transition is shown by the quick ejected prograde stars at $q \gtrsim 0.1$.  However, this does not seem to have an appreciable impact on the overall evolution of the binary.

The physical origin of the change in $K$ as a function of $q$ is most apparent in the behaviour of late-retrograde orbiters.
We emphasize, however, that the standard deviations associated to these mean values is usually much larger than the mean value.  Therefore, the stochastic nature of this result should definitely not be neglected when trying to draw conclusions. To assess the importance of the tails of the $(B-C)$ distribution, we focus on late-retrograde stars (i.e. those who show the unexpected trend). 

In Fig.~\ref{fig:BmC_mean_TH}, we report how the mean value $\langle B-C\rangle$ changes by considering only simulations with $|B-C|$ below a certain threshold, i.e., $|B-C|<|B-C|_{\rm threshold}$. The comparison between Fig.~\ref{fig:C_B_mean} and Fig.~\ref{fig:BmC_mean_TH} shows that the mean value is mostly determined by moderately strong encounters with $10<|B-C|<10^3$ (except perhaps for the $q=10^{-3}$ where the high $|B-C|$ tail is also important), i.e., the overall behaviour is dictated more by the tails rather than the bulk of the distributions (events resulting in $|B-C|<10$ are the majority), making any attempt of explaining it in terms of simple analytical arguments quite challenging.

The circularisation or eccentricity growth critically depends on the balance between changes in energy and angular momentum. This is because the net value is evaluated through cancellations, such that all these quantities are strongly influenced by fluctuations. This can indeed explain why the late ejection sub-population, despite being outnumbered by its quick counterpart, is so influential in determining the global value of $K$. 

\section{Discussion and conclusions}
\label{sec:discussion}

By investigating the problem of the scattering of unbound stars against an eccentric MBHB as a function of the binary mass ratio $q$, we showed that below $q\approx10^{-3}$, the interactions tend to circularise the binary. The origin of this behaviour has been identified as being associated with a sudden change in the angular momentum exchange of a specific subset of interactions, i.e., stars approaching the MBHB on retrograde orbits, that become temporarily bound to the MBHB (late-retrograde encounters).

Because of the chaotic nature of the interaction, and because the eccentricity evolution is determined by a subtle cancellation between energy and angular momentum exchange (via the $B-C$ combination in equation \ref{eq:deltae}), double checking the robustness of this result is imperative. First and foremost, we found qualitatively and quantitatively consistent results upon comparing our integrations to those obtained by \citet{Rasskazov2019} using the high accuracy N-body code ARCHAIN \citep{Mikkola2008}. We checked that the bimodality in time distribution as well as the behaviour of the different sub-populations of scatterings defined here is the same in the scattering experiments employed for the R19 study. We then checked that results are insensitive to several technical aspects of the integration, e.g. tolerance of the integrator, precise radius at which the three-body interaction is switched-on, and so on. None of the tests performed hinted to a possible numerical artefact.

Both our runs and those in R19 integrated the full three-body problem using $m_1=10^6\msun$ and $m_*=1\msun$. It is therefore legit to ask whether there is also a dependency of the results on the absolute mass scale of the system. To this end, we ran a series of scattering experiments with $m_1=10^7\msun, 10^5\msun$ and $m_*=1\msun$ in the mass ratio range $3\times10^{-4}<q<3\times10^{-3}$, finding the same change in $\langle B-C\rangle$ for late-retrograde encounters around $q\approx 10^{-3}$. We thus conclude that the effect must be solely driven by the MBHB mass ratio. Another dependence that we do not explore here is that on the initial binary eccentricity, which we assumed to be $e=0.6$. Indeed, for this particular value, we witness the most extreme decreasing trend for $K(q)$ and $e=0.6$ providing the more promising testing ground to pin down the origin of this phenomenon. Since the effect is seen at all eccentricities \citep[see Fig. 2 of][]{Rasskazov2019}, it is hard to imagine that it has a different origin at different values of $e$. We are currently planning to extend the study to different eccentricities, which we defer to a future paper.

\begin{figure}
    \centering
    \includegraphics[width=0.47\textwidth]{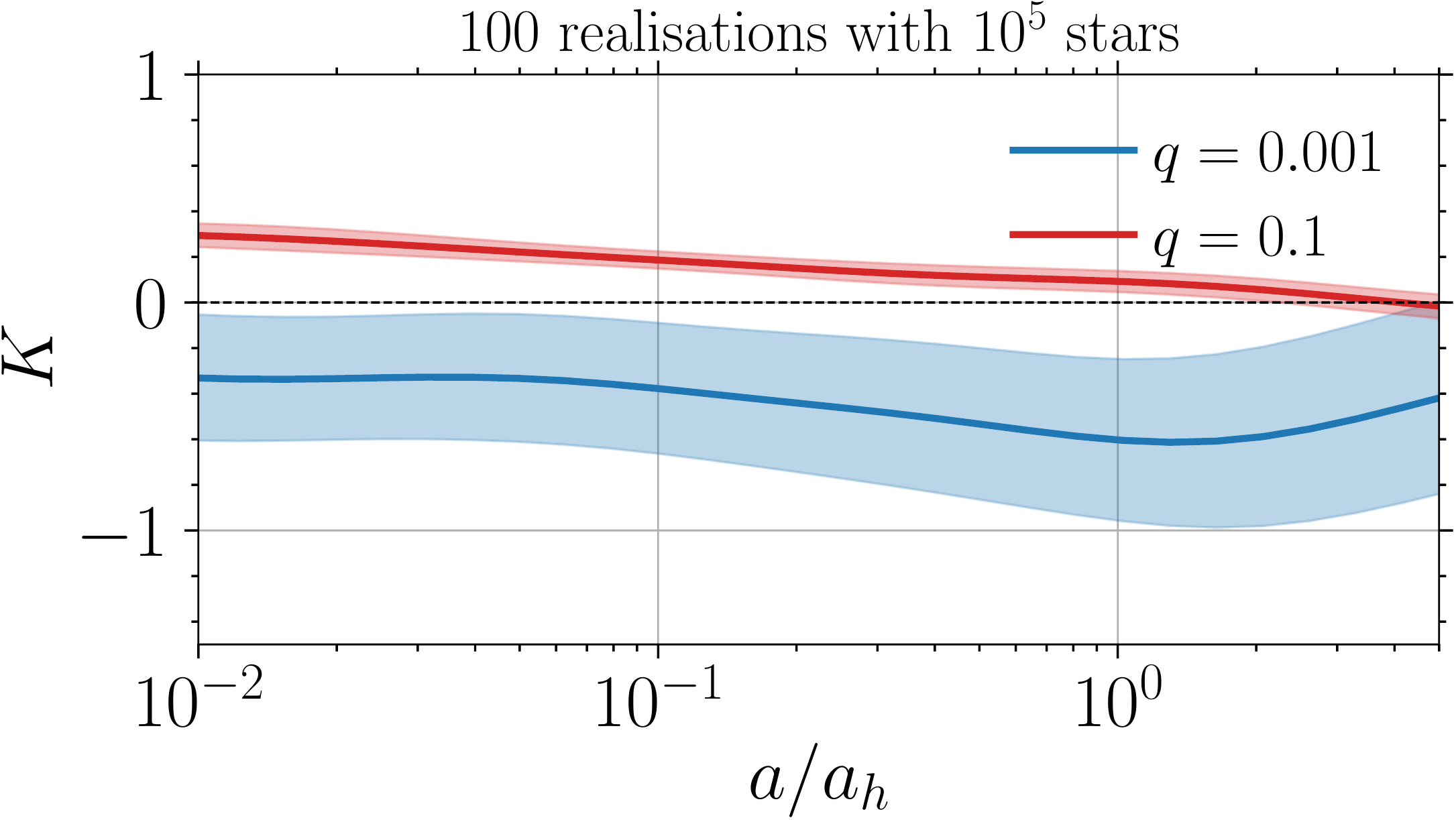}
    \caption{Eccentricity growth rate evaluated from 100 sub-samples of $10^5$ stars randomly extracted from the pool of $4\times 10^6$ stars for each $q$. Solid lines represent the average value over 100 realisations, while shaded areas denote 1-$\sigma$ dispersion (i.e. 68\% confidence region).}
    \label{fig:K100Re}
\end{figure}

Before drawing any astrophysical conclusions from these results, two important aspects of the problem should be borne in mind. First, using $H$ and $K$ to evolve the MBHB implicitly assumes that each interaction is independent. However, Fig.~\ref{fig:Tdist} shows that in the small $q$ limit, late scatterings can be extremely long lasting. The vertical lines in the figure mark $\tilde{T}=10^8$yr in units of $T_{\rm bin}$ for systems with $m_1=10^6\msun$ at $a=a_h$. This is the typical evolution timescale of the binary. It is encouraging that the peak of late scatterings occurs at $T\ll\tilde{T}$, meaning that the MBHB is not expected to significantly evolve during the interaction. To enforce this point, we also checked that the binary changes in energy and angular momentum are actually dominated by strong encounters, rather than by weak long-lasting secular interactions \citep[however see e.g.][for situations where the secular regime is actually relevant]{Hamers2019}. In fact, limiting our analysis to the sample of simulations with pericentre passage satisfying $r_p/a \leq 2$ (thus excluding the larger pericentre interval $r_p/a = [3,5]$) reveals that our results are substantially unchanged, thus indicating that most of the binary evolution is driven by close stronger encounters. It is, in any case, expected that several stars will be bound to the MBHB at any time. Although star-star interactions should be negligible, an N-body integration is ultimately warranted to confirm the full validity of the three body approximation.

Second, the chaotic nature of the interaction implies that the variance in the determination of $K$ can be much larger than its mean value. This appears to be particularly true for small mass ratios. In fact, as $q$ drops, rare strong close encounters with $m_2$ gain more weight in terms of energy and angular momentum exchange. 
It is interesting to estimate what level of stochasticity the evolution has for this relatively small number of ejections. For instance, let us consider a potential SgrA$^{*}$ companion with $q=10^{-3}$, which would have only $m_2\approx 4\times10^3\msun$. Extrapolating estimates given in figure 5 of  \cite{2005LRR.....8....8M} the binary would shrink to the GW emission regime by interacting with a mass in stars of $\approx 30\times m_2$, i.e., $\approx10^5$ stars. To this end, we compute $H$ and $K$ by generating 100 sub-samples of $10^5$ interactions randomly extracted form the pool of $4\times 10^6$. The result of this exercise is displayed in Fig.~\ref{fig:K100Re} for $q=0.001$ and $q=0.1$, which highlights the much larger dispersion in the former case, as mentioned above. The influence of stochasticity at low mass ratios is clearly appreciable. In fact, $K$ shows a remarkably high dispersion, consistent with zero at the 1-$\sigma$ level. It is worth noticing, however, that $10^5$ {\it scattering experiments} do not directly correspond to $10^5$ interactions. In fact $K$ is evaluated by weighting the experiments over a Maxwellian distribution, which gives different weights to different incoming velocities. Moreover, not all the interacting stars are eventually ejected (see e.g., Fig.~5 in R19). Nevertheless, this simple test shows that using averaged quantities like $K$ to evolve individual low-$q$ MBHBs is not justified and the only sensible statement that can be made is that three-body scatterings tend to circularise low $q$ MBHBs {\it on average}.\footnote{A more rigorous way to describe the dynamical evolution of an MBHB when stochasticity is important could be via the Fokker-Planck formalism \citep[see e.g.][]{Rasskazov2017}. We plan to address this point in a future follow-up study.}

Finally, we stress that circularisation causes a dramatic increase of the gravitational wave coalescence timescale. This might favour the accumulation of multiple IMBHs around a MBH, with potentially interesting dynamical consequences eventually affecting the global rate of such inspirals observable by the future Laser Interferometer Space Antenna \citep{as18}. Although we identified an association between late-retrograde encounters and the origin of low $q$ MBHB circularisation, we still lack a complete understanding of the underlying dynamical mechanism. We defer this to future work, in which we plan an in-depth study of individual loosely bound orbits.

\section*{Acknowledgements}

Numerical calculations have been made possible through a CINECA-INFN agreement, providing access to resources on GALILEO and MARCONI at CINECA. GF acknowledges support from a CIERA postdoctoral fellowship at Northwestern University. MAS acknowledges financial support by the Alexander von Humboldt Foundation, and support by the COST Action CA16104 ``GWVerse'' and the Deutsche Forschungsgemeinschaft (DFG, German Research Foundation) -- Project-ID 138713538 -- SFB 881 (``The Milky Way System''). AR is supported by the European Research Council (ERC) under the European Union's Horizon 2020 research and innovation program ERC-2014-STG under grant agreement No 638435 (GalNUC).



\bibliographystyle{mnras}
\bibliography{biblio} 




\bsp	
\label{lastpage}
\end{document}